\documentclass[preprint,twocolumn,5p]{elsarticle}
\usepackage{atlasphysics}
\usepackage{preprintcover}
\usepackage[hyperindex,breaklinks]{hyperref}
\biboptions{sort&compress}

\journal{Physics Letters B}

\PreprintCoverPaperTitle{Search for supersymmetry in events with photons, bottom quarks, and missing transverse momentum in proton-proton collisions at a centre-of-mass energy of 7~TeV with the ATLAS detector} 
\PreprintIdNumber{CERN-PH-EP-2012-308}
\PreprintCoverAbstract{A search has been performed for the experimental signature of an isolated photon with high transverse momentum, at least one jet identified as originating from a bottom quark, and high missing transverse momentum.
Such a final state may originate from supersymmetric models with gauge-mediated supersymmetry breaking in events in which one of a pair of higgsino-like neutralinos decays into a photon and a gravitino while the other decays into a Higgs boson and a gravitino.
The search is performed using the full dataset of 7 TeV proton-proton collisions recorded with the ATLAS detector at the LHC in 2011, corresponding to an integrated luminosity of 4.7 fb$^{-1}$.
A total of 7 candidate events are observed while 7.5~$\pm$~2.2 events are expected from the Standard Model background.
The results of the search are interpreted in the context of general gauge mediation to exclude certain regions of a benchmark plane for higgsino-like neutralino production.
}
\PreprintJournalName{Physics Letters B}

\begin{document}

\begin{frontmatter}

\title{Search for supersymmetry in events with photons, bottom quarks, and missing transverse momentum in proton-proton collisions at a centre-of-mass energy of 7~TeV with the ATLAS detector}
\author{The ATLAS Collaboration}

\begin{abstract}
A search has been performed for the experimental signature of an isolated photon with high transverse momentum, at least one jet identified as originating from a bottom quark, and high missing transverse momentum.
Such a final state may originate from supersymmetric models with gauge-mediated supersymmetry breaking in events in which one of a pair of higgsino-like neutralinos decays into a photon and a gravitino while the other decays into a Higgs boson and a gravitino.
The search is performed using the full dataset of 7 TeV proton-proton collisions recorded with the ATLAS detector at the LHC in 2011, corresponding to an integrated luminosity of 4.7 fb$^{-1}$.
A total of 7 candidate events are observed while 7.5~$\pm$~2.2 events are expected from the Standard Model background.
The results of the search are interpreted in the context of general gauge mediation to exclude certain regions of a benchmark plane for higgsino-like neutralino production.
\end{abstract}


\end{frontmatter}

\newcommand{\etogamma}{\ensuremath{e \rightarrow \gamma}}

\newcommand{\pho}{\phantom{0}}
\newcommand{\bslash}{\ensuremath{\backslash}}
\newcommand{\BibTeX}{{\sc Bib\TeX}}
\newcommand{\fb}{fb$^{-1}$}

\newcommand{\jetpt}{$p_{\rm{T,jet}}$}
\newcommand{\Wbb}{$Wb \bar b$}
\newcommand{\etmiss}{\ensuremath{E_{\rm{T}}^{\rm{miss}}}}
\newcommand{\bbgamma}{$b \bar b \gamma$}
\newcommand{\ttgamma}{$t \bar t \gamma$}
\newcommand{\mgl}{\ensuremath{m_{\tilde g}}}
\newcommand{\msq}{$m_{\tilde q}$}
\newcommand{\mnt}{\ensuremath{m_{\tilde{\chi}^0_1}}}
\newcommand{\hts}{$H_{\rm{T}}$}
\newcommand{\mts}{\ensuremath{m_{\rm{T}}(\gamma,E_{\rm{T}}^{\rm{miss}})}}
\newcommand{\neut}{\ensuremath{\tilde{\chi}^0_1}}
\newcommand{\ebf}{$E_{\rm{b_1}}$}
\newcommand{\ebs}{$E_{\rm{b_2}}$}
\newcommand{\ejf}{$E_{\rm{j_1}}$}
\newcommand{\ejs}{$E_{\rm{j_2}}$}

\section{Introduction}\label{sec:intro}

Theories of gauge-mediated supersymmetry breaking (GMSB) presume a hidden sector in which supersymmetry is broken and the symmetry breaking is communicated to the visible sectors through Standard Model gauge boson interactions~\cite{Dine:1981gu,AlvarezGaume:1981wy,Nappi:1982hm,Dine:1993yw,Dine:1994vc,Dine:1995ag}.
Such theories are attractive because the hypothesis of an intermediate hidden sector suppresses the magnitude of flavour-changing neutral currents.
The lightest supersymmetric particle (LSP) in GMSB is the ultra-light gravitino ($\tilde G$), which under certain circumstances is a viable dark matter candidate~\cite{Feng:2010gw}.
The next-to-lightest supersymmetric particle (NLSP) may be the lightest neutralino \neut, often assumed to be a bino-like particle.
The bino is the supersymmetric partner of the U(1) gauge field, coupling to the photon and $Z$ boson with strengths that are determined by the weak mixing angle.
This results in the \neut~decaying predominantly to a photon and the LSP.
The classical signature of GMSB is, therefore, events with two isolated energetic photons and large missing transverse momentum (\etmiss).
Searches for such a signature at the LHC and the Tevatron established strong experimental constraints on GMSB models~\cite{Abazov:2010us, Aad:2012ae}.
Recent extensions of the original GMSB idea, known as general gauge mediation (GGM)~\cite{Meade:2008wd}, evade these limits by allowing decoupled mass scales for strongly-interacting supersymmetric partners of the Standard Model particles.

In the GGM models considered in this Letter, the neutralino has higgsino or neutral wino (supersymmetric partners of the Higgs and neutral $W$ bosons) components instead of being predominantly bino-like, and therefore, in addition to its conventional decay to a gravitino and a photon, it may decay to a gravitino and a Higgs boson or to a gravitino and a $Z$ boson.
This GGM signature could be identified as an excess of events with pairs of neutralinos decaying to these bosons, in all combinations, associated with high \etmiss~\cite{Ruderman:2011vv}.
In particular, for a light Higgs boson ($m_h<130\,\mathrm{GeV}$), which decays predominantly to $b\bar b$, one final-state signature is the combination of an isolated high transverse momentum (\pt) photon, jets originating from bottom quarks, and high \etmiss.
Such a signature arises when one neutralino decays to a gravitino and a photon and the other to a gravitino and a Higgs boson.
This decay mode is therefore significant when both branching fractions are large, namely when the bino mass term $M_1$ approximately equals the higgsino mass parameter $-\mu$~\cite{Dine:1981gu}.

This Letter describes the search for events with a ``$\gamma$+$b$+\etmiss~topology'', consisting of an isolated high-\pt~photon, large \etmiss, and at least one jet that contains a $b$-hadron (``$b$-tagged jet''), in the full dataset of $\sqrt{s}=7\,\mathrm{TeV}$ $pp$ collisions recorded in 2011 with the ATLAS detector at the LHC, corresponding to a total integrated luminosity of $4.7\,\mathrm{fb}^{-1}$.
This signature is complementary to searches for diphoton production accompanied by \etmiss~\cite{Aad:2011zj, Chatrchyan:2011wc}, searches for $b$-jet production plus \etmiss~\cite{ATLAS:2012ah,Aad:2012pq}, searches for lepton production plus \etmiss~\cite{Aad:2012ms}, and searches for $Z$ bosons accompanied by photons and \etmiss~\cite{Abazov:2012ri}.
The $\gamma$+$b$+\etmiss~topology has not been studied in any previous search and therefore the present analysis can also be considered as a model-independent search for new phenomena in this final state.


\section{ATLAS detector} \label{sec:detector}
The ATLAS experiment~\cite{Aad:2008zzm} is a multi-purpose particle physics detector with a forward-backward symmetric cylindrical geometry and nearly $4\pi$ coverage in solid angle\footnote{ATLAS uses a right-handed coordinate system with its origin at the nominal interaction point (IP) in the centre of the detector and the $z$-axis along the beam pipe. The $x$-axis points from the IP to the centre of the LHC ring, and the $y$-axis points upward. Cylindrical coordinates ($r$, $\phi$) are used in the transverse plane, $\phi$ being the azimuthal angle around the beam pipe. The pseudorapidity is defined in terms of the polar angle $\theta$ as $\eta = -\ln \tan(\theta$/2).}.
The collision point is surrounded by inner tracking devices followed by a superconducting solenoid providing a 2~T magnetic field, a calorimeter system, and a muon spectrometer.
The inner tracker provides precision tracking of charged particles for pseudorapidities $|\eta| < 2.5$.
It consists of pixel and silicon microstrip detectors inside the transition radiation tracker.
The calorimeter system has liquid argon (LAr) or scintillator tiles as the active media.
In the pseudorapidity region $|\eta| < 3.2$, high-granularity LAr electromagnetic (EM) sampling calorimeters are used.
An iron/scintillator tile calorimeter provides hadronic coverage for $|\eta| < 1.7$.
The end-cap and forward regions, spanning $1.5<|\eta| <4.9$, are instrumented with LAr calorimeters for both EM and hadronic measurements.
The muon spectrometer consists of three large superconducting toroids with 24 coils, a system of trigger chambers, and precision tracking chambers, which provide triggering and tracking capabilities in the ranges $|\eta| < 2.4$ and $|\eta| < 2.7$, respectively.


\section{Simulated samples} \label{sec:simulations}
Standard Model processes that constitute the background to this search are simulated using several different generator programs.
Events with single- or pair-production of top quarks are simulated using the MC@NLO~\cite{Frixione:2002ik} generator with the CT10~\cite{Lai:2010vv} parton distribution functions (PDFs), where the generator is interfaced to the HERWIG~\cite{Corcella:2000bw} and JIMMY~\cite{Butterworth:1996zw} programs to include effects of fragmentation and hadronization and the underlying event.
The POWHEG generator~~\cite{Nason:2004rx,Frixione:2007vw,Alioli:2010xd} is also used for studies of systematic uncertainties in these events.
The $t\bar{t}\gamma$ background is simulated with the WHIZARD~\cite{Kilian:2007gr} generator, which incorporates a full calculation of the seven-particle final states $\ell\nu_{\ell} q\bar{q}'b\bar{b}\gamma$ and $\ell\bar\nu_{\ell} \bar\ell'\nu_{\ell'} b\bar{b}\gamma$ (with $\ell/\ell' = e, \mu, \tau$) at leading order (LO).
These events are generated with the CTEQ6L1~\cite{Pumplin:2002vw} PDFs and hadronized with HERWIG; additional photon(s) that may be radiated in the fragmentation process are generated by PHOTOS~\cite{Barberio:1990ms}.
Multijet background (``QCD multijet'') events are simulated using the PYTHIA~\cite{Sjostrand:2006za} generator. Diboson background events ($W^+W^-$, $W^{\pm}Z$, and $ZZ$) are simulated using HERWIG. Events with vector bosons accompanied by $b\bar{b}$ or light jets are simulated using ALPGEN~\cite{Mangano:2002ea} and HERWIG~\cite{Corcella:2000bw}.

The production of signal events is simulated in two separate two-dimensional benchmark grids of points defined by specific GGM model parameters.
The first grid has various gluino and neutralino masses (\mgl,\mnt), while the second grid has varying squark and neutralino masses (\msq,\mnt).
The fundamental parameters $M_{1}$ and $\mu$ together determine the lightest neutralino mass and are adjusted in such a way that the following branching ratios of the \neut~ are approximately constant: BR$(\neut\rightarrow h+\tilde{G}) \approx 56\%$, BR$(\neut\rightarrow\gamma+\tilde{G}) \approx 33\%$, and BR$(\neut\rightarrow Z+\tilde{G}) \approx 11\%$.
These numbers vary by $\pm 2\%$ throughout the grids.
The value of $\mu$ is chosen to be negative in order to make the branching ratio of the \neut~to the lightest Higgs boson greater than that to the $Z$ boson.
Masses of the sleptons and coloured supersymmetric particles not used to make the grid are set to 2.5 TeV, and the lightest Higgs boson is in the decoupled regime with $m_{A} = 2\,\mathrm{TeV}$ and $m_{h}= 115\,\mathrm{GeV}$, which results in a branching ratio $\mathrm{BR}(h \rightarrow b\bar b)=74\%$.
Other parameters are the wino mass $M_{2} = 2.5\,\mathrm{TeV}$, the ratio of Higgs doublet vacuum expectation values $\tan\beta = 1.5$, and the neutralino decay length $c\tau < 0.25\,\mathrm{mm}$.
The small effect of a different choice of Higgs boson mass, $m_h = 125\,\mathrm{GeV}$, is discussed in Sec.~\ref{sec:results}.
More generally, different choices of these parameters can modify slightly the relevant branching ratios but do not affect significantly the overall sensitivity reach for models of gauge mediation.
The full mass spectrum and decay widths are calculated using SUSPECT, SDECAY, and HDECAY with the SUSY-HIT interface~\cite{Djouadi:2006bz}. Events are generated with Herwig++~\cite{Bahr:2008pv}.

The signal production rate is dominated at high neutralino masses by strong production of gluinos and squarks, but at low neutralino masses the direct production of charginos and neutralinos is greatly enhanced.
Signal cross sections are calculated to next-to-leading order in the strong coupling constant (NLO) using PROSPINO2~\cite{Beenakker:1996ch}\footnote{The addition of the resummation of soft gluon emission at next-to-leading-logarithmic accuracy (NLL)~\cite{Beenakker:1996ch,Kulesza:2008jb,Kulesza:2009kq,Beenakker:2009ha,Beenakker:2011fu} is performed in the case of strong SUSY pair-production.}.
The nominal cross section and its uncertainty are taken from an envelope of cross-section predictions using different PDF sets and factorization and renormalization scales, as described in Ref.~\cite{Kramer:2012bx}.
The PDF sets used for those calculations are CTEQ6.6M~\cite{Nadolsky:2008zw} and MSTW2008NLO~\cite{Martin:2009iq}.

Monte Carlo simulated event samples are generated with multiple $pp$ interactions (pile-up) and are re-weighted by matching the distribution of the number of interactions per bunch crossing to that observed in the data.
The samples are passed through the GEANT4~\cite{Agostinelli:2002hh,Aad:2010ah} simulation of the ATLAS detector and the same reconstruction software used for the data.


\section{Event reconstruction} \label{sec:objects}

Jets are reconstructed from calibrated clustered energy deposits in the calorimeter using the anti-$k_t$ jet clustering algorithm~\cite{Cacciari:2008gp} with radius parameter $R = 0.4$.
Clusters of calorimeter cells are seeded by cells with energy significantly above the measured noise.
Jet energies are corrected for the effects of calorimeter non-compensation and inhomogeneities using \pt- and $\eta$-dependent calibration factors based on Monte Carlo simulations validated with extensive test-beam and collision-data studies~\cite{Aad:2011he}.
Reconstructed jets with $\pt>20\,\mathrm{GeV}$ and $|\eta| < 2.8$ are used in this analysis.

A multivariate $b$-tagging algorithm that exploits both impact parameter and secondary vertex information is used to identify jets with $|\eta| < 2.5$ containing a $b$-hadron~\cite{ATLAS-CONF-2012-043}.
 The working point used in this analysis has a 60\% efficiency on a sample of $b$-jets from simulated $t\bar t$ events, with typical misidentification rates of 12\% for $c$-jets and less than 0.2\% for light-quark/gluon jets with $\pt > 20\,\mathrm{GeV}$ and $|\eta | <2.5$.

A photon candidate must have transverse momentum $p_\mathrm{T} > 20\,\mathrm{GeV}$ and must fulfil a set of ``tight'' identification requirements~\cite{Aad:2010sp}.
Moreover, the cluster associated with the photon should have $|\eta| < 2.37$ and should not be in the transition region between the barrel and end-cap calorimeters ($1.37 < |\eta | < 1.52$).
An isolation criterion is applied in order to suppress the background from photons originating inside jets: the total calorimeter energy deposit, not including the photon candidate, inside a cone of $\Delta R= \sqrt{(\Delta \eta)^2 + (\Delta \phi)^2} =0.2$ around the photon direction is required to be less than 5~GeV.
Photon candidates identified from conversions are included, but, in order to suppress the background from primary electrons misidentified as photons, the tracks of converted photon candidates are required to have no hits in the pixel detector.

Electron candidates are clustered energy deposits in the electromagnetic calorimeter matched to a track in the inner detector.
They are required to have $\pt>20\,\mathrm{GeV}$ and $|\eta| <2.47$, and must satisfy the ``medium'' electron shower shape and track selection criteria described in Ref.~\cite{Aad:2011mk}.
As for photons, electron candidates in the calorimeter transition region are vetoed.

Muon candidates with $|\eta| <2.4$, reconstructed by combining tracks in the inner detector and tracks in the muon spectrometer, are required to have $\pt>10\,\mathrm{GeV}$ and also to pass muon quality requirements~\cite{Aad:2011cwa}.

The measurement of the missing transverse momentum, including its magnitude \etmiss, is based on the vector sum of the reconstructed transverse momenta in the event.
Objects included in the sum are muons and electrons with $\pt>10\,\mathrm{GeV}$, photons with $\pt>20\,\mathrm{GeV}$, jets with $\pt>20\,\mathrm{GeV}$, and calibrated calorimeter clusters that are not associated with any object with $|\eta| <4.9$, as described in Ref.~\cite{Aad:2012re}.

Any jet candidate lying within a distance $\Delta R < 0.2$ from an electron or photon is discarded.
Also, in order to ensure that selected leptons and photons are not purely the result of hadronic activity, electrons and photons with distances $0.2 < \Delta R < 0.4$ from a jet are rejected, as are muons within $\Delta R < 0.4$ of a jet. 
The difference in requirements reflects the fact that only photons and electrons can potentially be reconstructed as jets.
Since one of the main backgrounds in this analysis is due to electrons misidentified as photons, a preliminary suppression of the background is achieved by labelling an object an electron whenever an electron/photon ambiguity exists and by discarding the photon candidate if it lies within $\Delta R < 0.2$ of any electron.


\section{Event selection}\label{sec:selection}
The data sample is collected with a trigger requiring at least one photon passing ``loose'' identification requirements~\cite{Aad:2010sp} with $\pt>80\,\mathrm{GeV}$; this trigger is fully efficient for the selection described below.
The following selection criteria were optimized to maximize the sensitivity to the GGM scenarios considered, especially gluino/squark production: a candidate event should contain a photon with $\pt>125\,\mathrm{GeV}$, at least two jets with $p_{\rm{T}}>20\,\mathrm{GeV}$, at least one of which is $b$-tagged, and $\etmiss > 150\,\mathrm{GeV}$.
The transverse mass of the photon and the missing transverse momentum $\mts = \sqrt{2 \etmiss p_{\mathrm{T}}^{\gamma} (1-\cos\phi)}$, where $\phi$ is the azimuthal angle between the missing transverse momentum and the photon, is required to be greater than $100\,\mathrm{GeV}$.
This criterion removes events in which electrons or decay products of $\tau$ leptons, originating from $W$ decay, are misidentified as photons.
The minimum azimuthal angle between the \etmiss~direction and each of the two leading jets must be greater than 0.4.
This condition suppresses multijet events in which the measured \etmiss~is due mostly to jet mismeasurement effects.
Events with an identified electron or muon satisfying the criteria given in Section~\ref{sec:objects} are vetoed.
This veto suppresses dileptonic and semileptonic \ttbar~events with a prompt photon or with a jet misidentified as a photon, and dileptonic events with an electron or a $\tau$ lepton misidentified as a photon.
Finally, events with a second photon with $\pt>50\,\mathrm{GeV}$ are rejected.
The main selection requirements are summarized in Table~\ref{table:selection}.

\begin{table}[!htbp]
\begin{center}
\begin{tabular}{c|c}
\hline
1 photon ($p_{\rm{T}}>125\,\mathrm{GeV}$) & $\mts>100\,\mathrm{GeV}$ \tabularnewline
$\geq 2$ jets ($p_{\rm{T}} > 20\,\mathrm{GeV}$) & $\Delta \phi$ (\etmiss, jet) $> 0.4$ \tabularnewline
$\geq 1$ $b$-tagged jet & veto $e/\mu$ \tabularnewline
$\etmiss > 150\,\mathrm{GeV}$  & veto second photon \tabularnewline
\hline
\end{tabular}
  \caption{Summary of event selection requirements.}
 \label{table:selection}
\end{center}
\end{table}


\section{Background estimation}\label{sec:bg}
Events from $\ttbar$ production with a $W$ boson decaying into leptons in the final state (leptonic \ttbar~background) contain a pair of $b$-jets and genuine \etmiss.
These events may survive the signal selection procedure if an isolated high-\pt~photon candidate is also present.
Such a photon may be the result of the misidentification of an electron produced in the leptonic $W$ decay, a genuine prompt photon, or a $\tau$ decay product or jet misidentified as a photon.
All processes that give rise to final states $W(\rarrow \ell\nu)+X$, including leptonic \ttbar, diboson, and single top backgrounds, are estimated using data-driven methods. 
Another large background estimated with data-driven methods is from multijet events.
Finally, the small contribution from $Z(\rarrow\nu\bar{\nu}$)+jets background is estimated using Monte Carlo simulation.

A control sample (CS) is defined by selecting events according to the criteria described in Section~\ref{sec:selection} but replacing the photon selection by requiring the presence of an electron.
Once the probability of an electron being misidentified as a photon (the ``\etogamma~misidentification rate'') is known, the number of events in the signal region with misidentified electrons can be deduced from this CS.
The \etogamma~misidentification rate for different $\eta$ regions is measured by selecting events with a photon and an electron in which the $\gamma e$ invariant mass is less than $20\,\mathrm{GeV}$ from the nominal $Z$ boson mass of $91.2\,\mathrm{GeV}$.
The electron is required to pass the ``tight'' identification criteria~\cite{Aad:2010sp}, and the photon is required to pass the quality requirements of the signal region.
The number of $\gamma e$ events is then divided by the number of $e^+e^-$ pairs with one tight and one medium electron, and the ratio is taken to be the misidentification rate.
The average misidentification rate for photons with $p_{\rm{T}}>100\,\mathrm{GeV}$ is 1.8\%.
When this technique is applied to the data, $1.1\pm 0.1\,\mathrm{(stat.)}$ background events with electrons misidentified as photons are predicted in the signal region.

The prompt photon background cannot be separated from the backgrounds in which a jet or $\tau$ lepton is misidentified as a photon.
Therefore, a single CS is used to estimate these backgrounds.
The ``lepton control region'' is defined by requiring the presence of a lepton, in addition to the photon, and relaxing the \etmiss~cut to $80 < \etmiss < 150\,\mathrm{GeV}$ while keeping all the other selection criteria of Section~\ref{sec:selection}.
The lepton requirement strongly suppresses the multijet contamination, making it possible to use a lower \etmiss~threshold in order to increase the number of selected events and hence reduce the uncertainty on the background estimate.
The lower \etmiss~threshold is chosen to be $80\,\mathrm{GeV}$ to ensure that the \ttbar~background remains the dominant contribution in the lepton control region.
The results of the method for the signal region and the lepton control region are shown in Table~\ref{table:prompt_cs_subtract}.
In order to prevent double counting, the background with electrons misidentified as photons is subtracted, leaving 10.1 events in the CS.
Multiplying the 10.1 events observed in the CS by the simulation-based scale factor of $2.8/8.4 = 0.33$ gives a prediction of $3.4 \pm 1.7$~(stat.) prompt photon and misidentified jet/$\tau$ background events in the signal region.
The uncertainty is dominated by the limited number of events in the CS data.

An important issue in evaluating the scale factor with simulated events is that the MC@NLO generator does not produce \ttgamma~final states with a matrix element calculation; rather, it produces the \ttbar~hard process, and supplemental photon radiation is generated by HERWIG and PHOTOS.
The WHIZARD generator is better suited for \ttgamma~studies with high-$p_{\rm{T}}$ photons, since the photon is generated from a matrix element calculation.
To avoid double counting in the two samples, events in the MC@NLO simulation sample with a prompt photon are removed.
Even though the CS is dominated by \ttgamma~events, and the \ttgamma~simulations alone are used for the scale factor calculation, this technique gives a total estimate for all of the $W(\rarrow \ell\nu)+X$ background processes, which are present by construction in the CS.

\begin{table}[!htbp]
\begin{center}
\begin{tabular}{ccc}
  \hline
Sample  & Signal region & Lepton control region \tabularnewline
\hline
\ttbar~MC@NLO & 0.3 $\pm$ 0.2 & 0.5 $\pm$ 0.3 \tabularnewline
\ttbar$\gamma$ WHIZARD & 2.5 $\pm$ 0.2 & 7.9 $\pm$ 0.4 \tabularnewline
\hline
Total & 2.8 $\pm$ 0.3 & 8.4 $\pm$ 0.5 \tabularnewline
\hline
Data &   & 10.1 $\pm$ 3.5  \tabularnewline

\hline
  \end{tabular}
  \caption{Number of events in the signal region and in the lepton control region, as predicted by the \ttbar~MC@NLO and \ttgamma~WHIZARD calculations, after subtracting the overlapping contribution from electrons misidentified as photons.  Only statistical uncertainties are quoted.}
 \label{table:prompt_cs_subtract}
\end{center}
\end{table}

\begin{figure}[!htbp]
\begin{center}
\includegraphics*[width=0.49\textwidth]{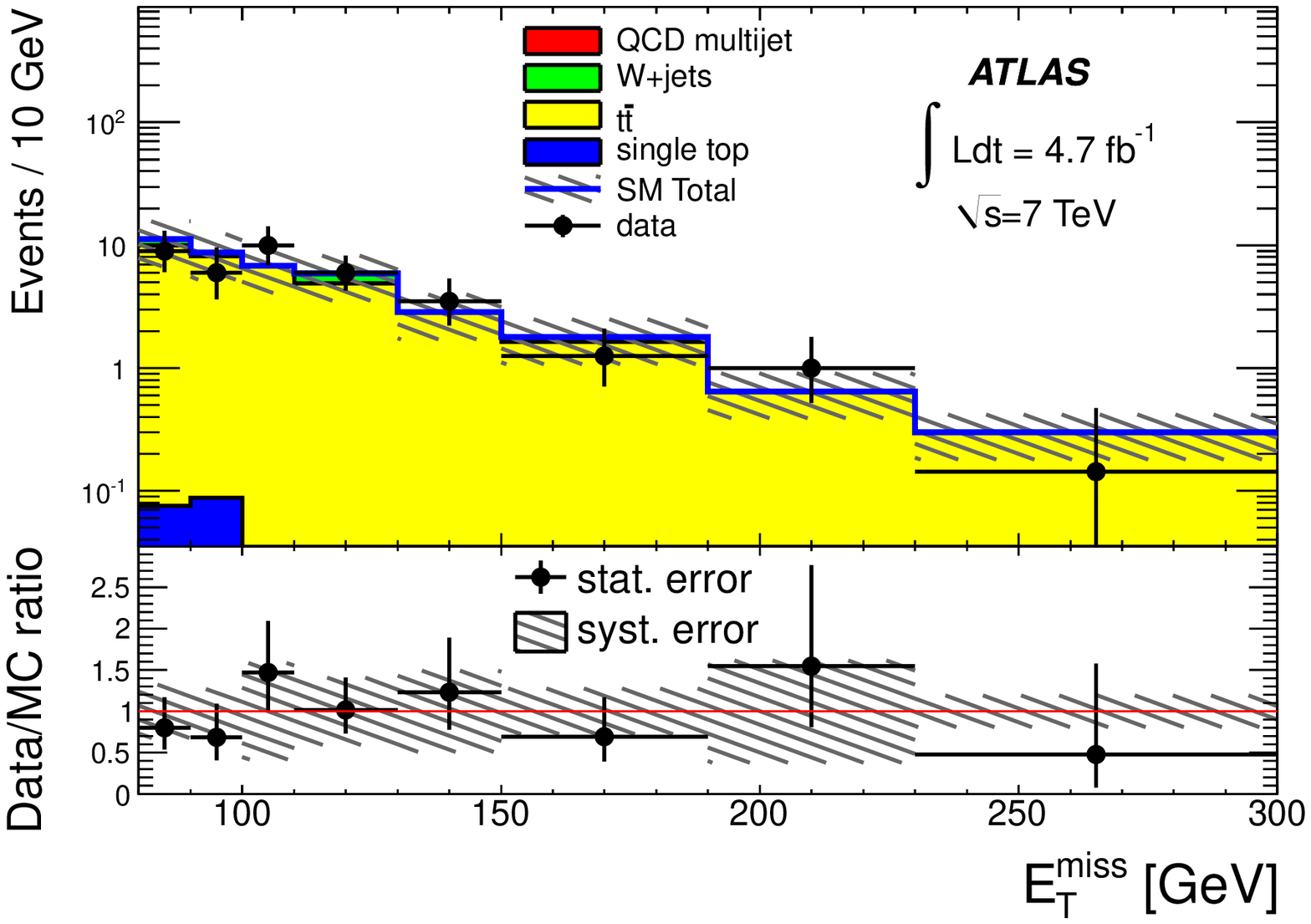}
\includegraphics*[width=0.49\textwidth]{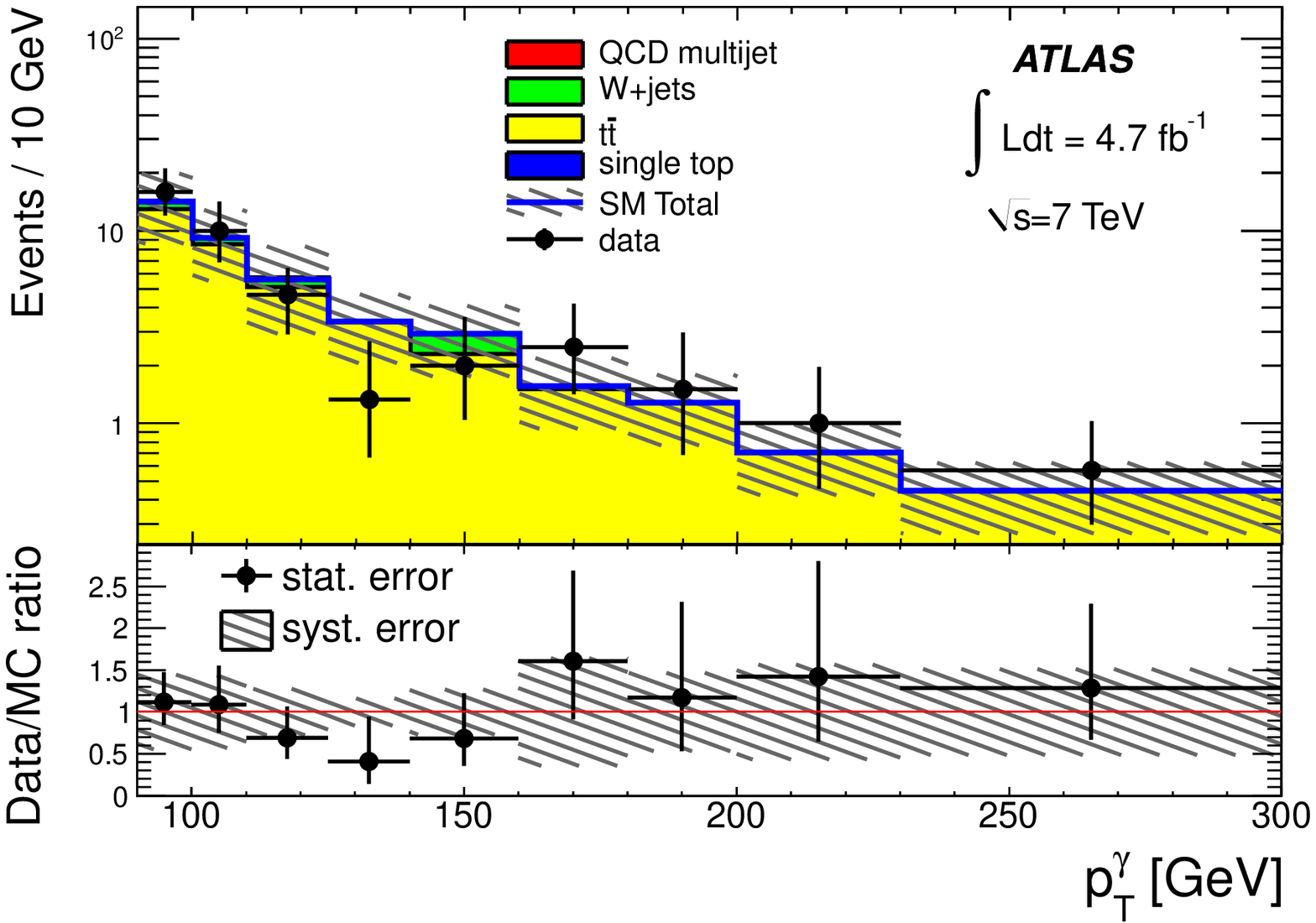}
\end{center}
\caption{The $E_{\rm{T}}^{\rm{miss}}$ spectrum in the lepton control region with a relaxed $p_{\rm{T}}^{\gamma}$ requirement of  $90\,\mathrm{GeV}$ and no \mts~cut (top) and the $p_{\rm{T}}^{\gamma}$ spectrum in the lepton control region with no upper bound on $E_{\rm{T}}^{\rm{miss}}$ and no \mts~cut (bottom).}
\label{fig:MET_ptgamma_MC_Data_tt_cr}
\end{figure}

To verify that the event characteristics used in this method are well modelled in the lepton control region,
the $E_{\rm{T}}^{\rm{miss}}$ and $p_{\rm{T}}^{\gamma}$ distributions in the data and simulation are shown in Fig.~\ref{fig:MET_ptgamma_MC_Data_tt_cr}.
The distributions agree within uncertainties.

Multijet events, another source of background, may contain genuine photons or misidentified jets that hadronize to an isolated $\pi^{0}$.
High \etmiss~is rare in multijet events but can be realized by mismeasured jets or by heavy-flavour quark jets decaying semileptonically.
To estimate the multijet contribution in the signal region (SR), control regions (CRs) are defined with events that fail the $b$-tag requirement or the \etmiss~requirement (see Table~\ref{table:qcd_simulations}).

\begin{table}[htbp]
\begin{center}
\begin{tabular}{ccc}
\hline
			&	\etmiss$<$100 GeV&	\etmiss~$>$~150 GeV  	\\\hline                       	
$\geq$ 1 $b$-tag 	&	CR1   			&       SR					\\
0 $b$-tag           	&	CR2    			&	CR3					\\
\hline
\end{tabular}
\caption{Definition of the control regions $\mathrm{CR1}, \mathrm{CR2}, \mathrm{CR3}$, and signal region SR for the multijet background estimation.}
\label{table:qcd_simulations}
\end{center}
\end{table}

The CR3 data sample is contaminated by \ttbar, single top, and $W/Z$+jets events that have genuine \etmiss, and this contamination must be removed.
This contamination $N_\mathrm{CR3}^{\mathrm{MC,multijet}}$ is estimated from the Monte Carlo simulation and accounts for approximately 45\% of the events in the CR3.
A scale factor between the tagged and untagged samples is calculated in the low $\etmiss<100\,\mathrm{GeV}$ control regions ($N_\mathrm{CR1}^{\mathrm{Data}}/N_\mathrm{CR2}^{\mathrm{Data}}$), and this scale factor is subsequently applied to the high-\etmiss~region of the untagged CS to obtain the prediction for the signal region:

\begin{eqnarray}
N_\mathrm{SR}^{\mathrm{Pred}} = \left(N_\mathrm{CR3}^{\mathrm{Data}}-N_\mathrm{CR3}^{\mathrm{MC,multijet}}\right)\times\left[\frac{N_\mathrm{CR1}^{\mathrm{Data}}}{N_\mathrm{CR2}^{\mathrm{Data}}}\right] \nonumber
\label{eqn:qcd}
\end{eqnarray}

In order to check the accuracy of this method, the background estimate is calculated after all selection requirements and then repeated without the \mts~$>$~100~GeV requirement, which is expected to have little effect on the multijet background.
The two calculations yield predictions of 3.3 $\pm$ 0.7 (stat.) $\pm$ 0.6 (syst.) events before the \mts~requirement and 2.7 $\pm$ 0.7 (stat.) $\pm$ 0.7 (syst.) events after all requirements, with uncertainty due only to limited statistics in the CRs.
The difference of 0.6 events is used as a systematic uncertainty associated with this method.
The number of expected QCD multijet events in the signal region is therefore 2.7 $\pm$ 1.1 events.

Finally, the $Z(\rarrow\nu\bar{\nu}$)+jets process is estimated, from studies of simulated events, to contribute 0.3 $\pm$ 0.3 events in the signal region.
The background from other sources is estimated to be negligible.


\section{Systematic uncertainties on the background}\label{sec:bg_uncert}
The main source of systematic uncertainties on the background is the scale factor derived from simulation for prompt photon and misidentified jet or $\tau$ processes.
The uncertainty on this factor is dominated by the theoretical uncertainties on the $\ttbar$ processes.
Uncertainties such as Monte Carlo modelling and different initial- and final-state radiation models are evaluated by comparing AcerMC (LO)~\cite{Kersevan:2004yg}, MC@NLO (NLO), and POWHEG (NLO) \ttbar~simulations.
The impact of using different fragmentation and hadronization models is estimated by comparing two POWHEG samples, one showered with HERWIG and the other with PYTHIA.
The uncertainty is defined as the greatest difference among the resulting scale factors with respect to the MC@NLO factor and is evaluated to be 17\%.
Other systematic uncertainties are smaller since the scale factor is a ratio of the event population in the signal and control regions and most of the uncertainties cancel out.
The effects of the jet energy scale~\cite{Aad:2012ag} and jet energy resolution uncertainties~\cite{Aad:2011he} are determined to be 4\% and 2\%, respectively, and the relative uncertainty due to the $b$-tagging efficiency is evaluated to be 1$\%$.
The systematic uncertainty in the photon identification is based on the results of data-driven measurements with $Z \rarrow e^+e^-$ decays and contributes 1$\%$ uncertainty in the scale factor.
The systematic uncertainty due to pile-up is estimated to cause background variations of up to 4$\%$, while the systematic uncertainty due to lepton identification, specifically in the lepton veto in the event selection, is estimated to be 6$\%$.
The impact of the luminosity uncertainty is less than $1\%$ because only the small contribution from $Z(\rarrow\nu\bar{\nu}$)+jets background is normalized using the integrated luminosity.


\section{Signal efficiencies and systematic uncertainties}\label{sec:signal_uncert}

The combined product of acceptance and efficiency of the event selection is calculated with simulated events for each point in the GGM benchmark grids.
Low $\mnt$ values typically result in gravitinos with relatively low \pt, which translates to lower efficiency for the \etmiss~requirement relative to high-$\mnt$ points.
A typical efficiency for high-mass gluino points ($\mgl=900\,\mathrm{GeV},\mnt=450\,\mathrm{GeV}$) is 10\%, including the branching ratio for all Higgs boson decays and the contribution from neutralino decays to $Z$ bosons that subsequently decay to $b\bar b$.
Uncertainties on the signal cross section due to PDFs, renormalization and factorization scales, and the strong coupling constant $\alpha_s$ are calculated separately for each production process as described in Ref.~\cite{Kramer:2012bx} and combined into an overall uncertainty that varies significantly for different signal points.
Most of the signal points have a combined cross-section uncertainty of 2--5$\%$ but the total uncertainty can reach 50$\%$ for the points with very large gluino masses.
The uncertainties on the signal acceptance include an uncertainty ranging from 3\% to 16\% due to the limited number of simulated events at each benchmark grid point.
The uncertainty on the jet energy scale and jet energy resolution, $b$-tagging efficiency, photon and lepton identification, luminosity, and pile-up are evaluated as in Section~\ref{sec:bg_uncert}.
The uncertainties on the jet energy scale and jet energy resolution vary from 1\% to 10\% across the different signal points.
The relative uncertainty on the signal selection efficiency due to the uncertainty in the $b$-tagging efficiency varies between 1$\%$ and 16$\%$ throughout the signal grid.
The systematic uncertainty on the photon identification is less than 6$\%$.
The systematic uncertainty on lepton identification is 3$\%$.
Scaling the number of pile-up events in simulation gives rise to variations of up to 6\% throughout the grid.
The systematic uncertainty on luminosity is evaluated to be 4$\%$~\cite{Aad:2011dr,ATLAS-CONF-2011-116}.
All the sources of described background and signal systematic uncertainties are summarized in Table~\ref{tab:bg_sys}.

\begin{table}[!htbp]
\begin{center}
\begin{tabular}{ccc}
\hline
Source of uncertainty & Background & Signal \tabularnewline
  \hline
Lepton identification & 6\% & 3\% \tabularnewline
Jet energy scale & 4\% & 1\%  \tabularnewline
Jet energy resolution & 2\% & \hspace{-0.2cm}$<1\%$  \tabularnewline
Photon identification & 1\% & 6\%  \tabularnewline
$b$-tagging & 1\% & 4\% \tabularnewline
Pile-up & 4\% & 1\% \tabularnewline
Theoretical uncertainties & 17\%  & 9\% \tabularnewline
Monte Carlo statistics & --  & 3\% \tabularnewline
Luminosity & \hspace{-0.2cm}$<1\%$ & 4\% \tabularnewline
 \hline
  \end{tabular}
  \caption{Summary of relative systematic uncertainties on the numbers of background and signal events (for the representative signal point $\mgl=900\,\mathrm{GeV}, \mnt=450\,\mathrm{GeV}$). Theoretical uncertainties on the background originate from the Monte Carlo modelling and different initial- and final-state radiation models. Theoretical uncertainties on the signal cross section originate from the PDFs, renormalization and factorization scales, and $\alpha_s$.}
 \label{tab:bg_sys}
\end{center}
\end{table}


\section{Results}\label{sec:results}
\begin{figure}[!htbp]
\begin{center}
\includegraphics*[width=0.49\textwidth]{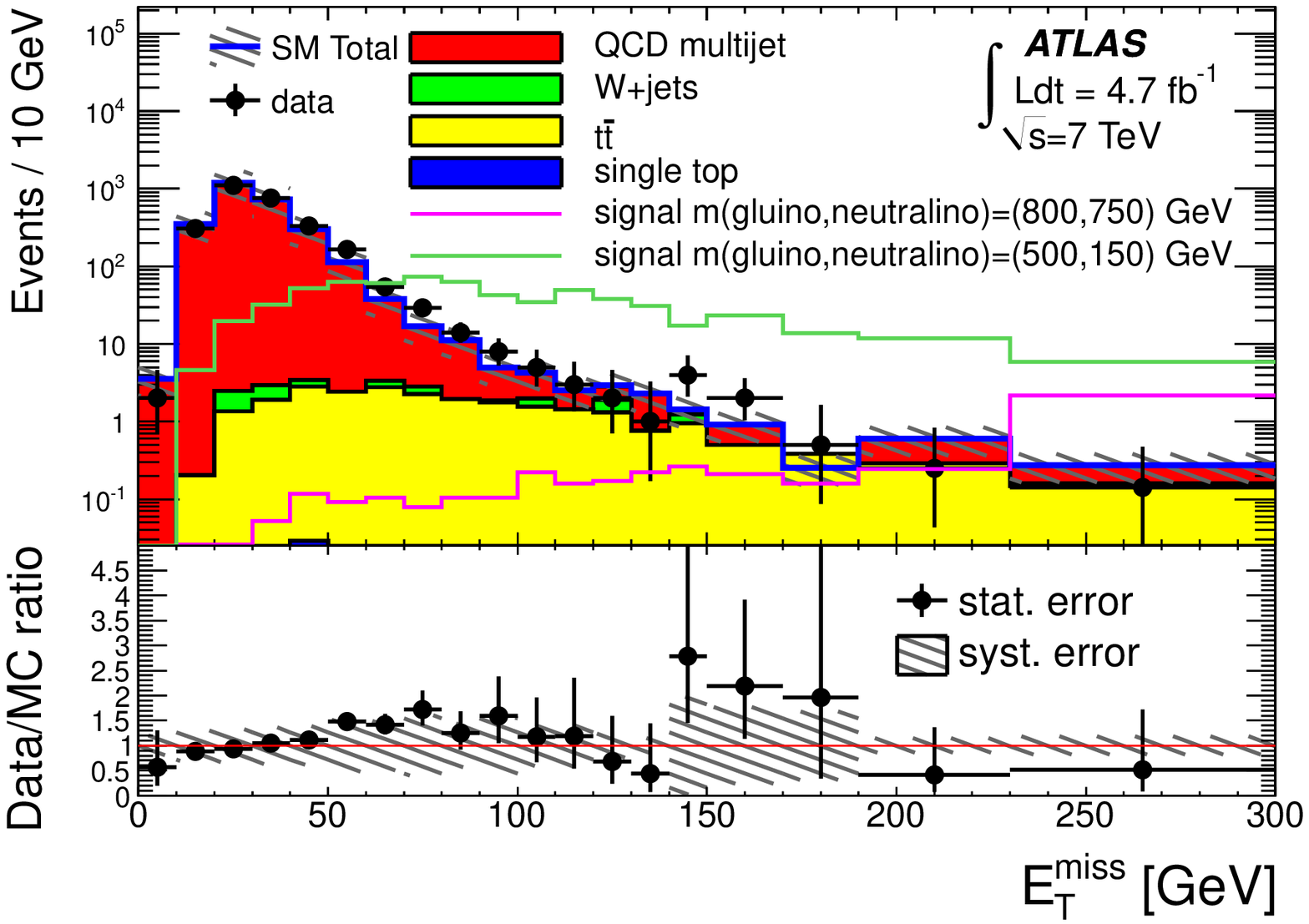}
\includegraphics*[width=0.49\textwidth]{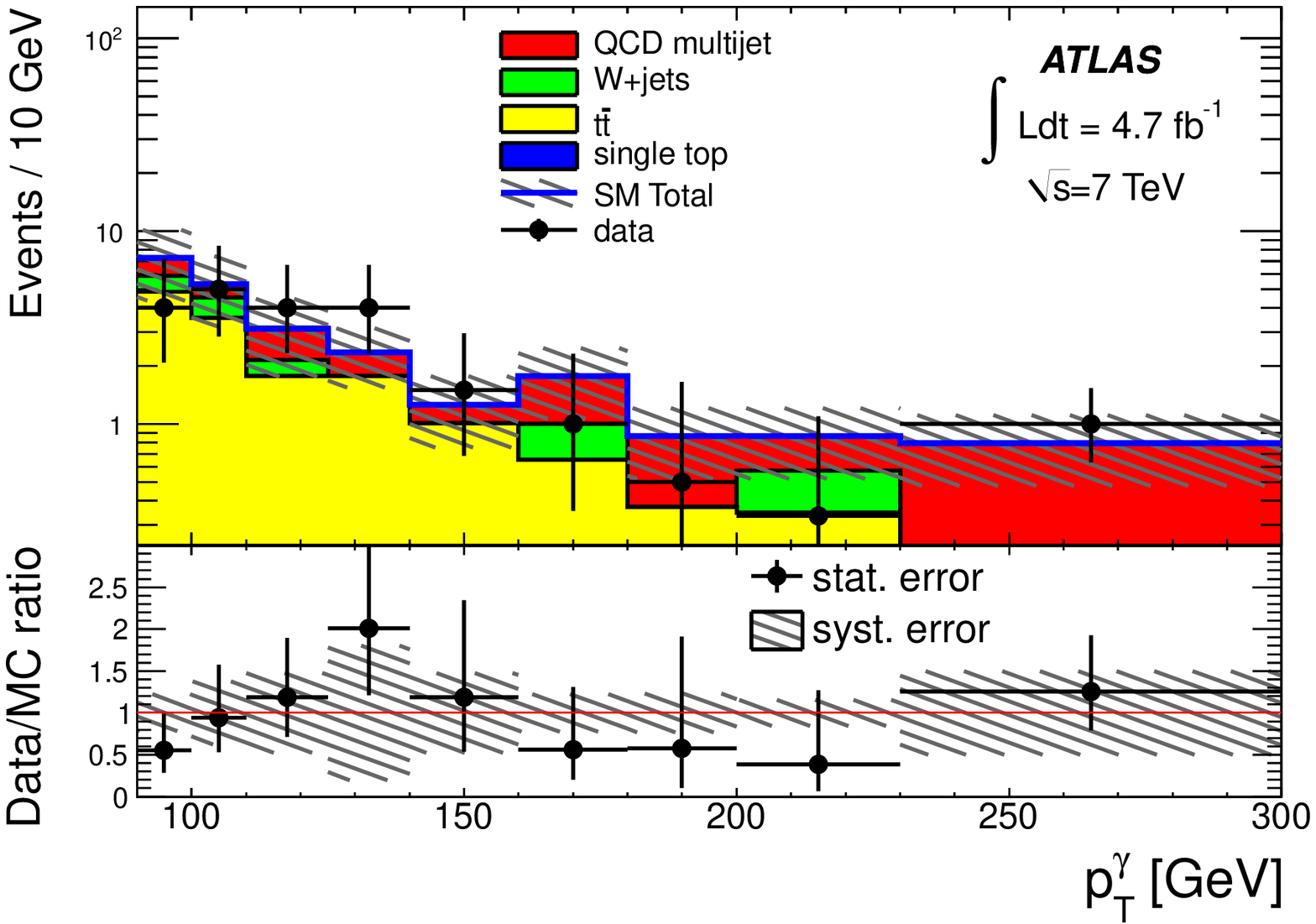}
\end{center}
\caption{The \etmiss~distribution after all selection criteria except the \etmiss~cut (top) and the $p_{\rm{T}}^{\gamma}$ distribution after all selection criteria except those on \mts~and $\Delta\phi(\etmiss,\mathrm{jet})$ (bottom).}
\label{fig:MET_ptgamma_MC_Data}
\end{figure}

\begin{figure}[!htbp]
\begin{center}
\includegraphics*[width=0.49\textwidth]{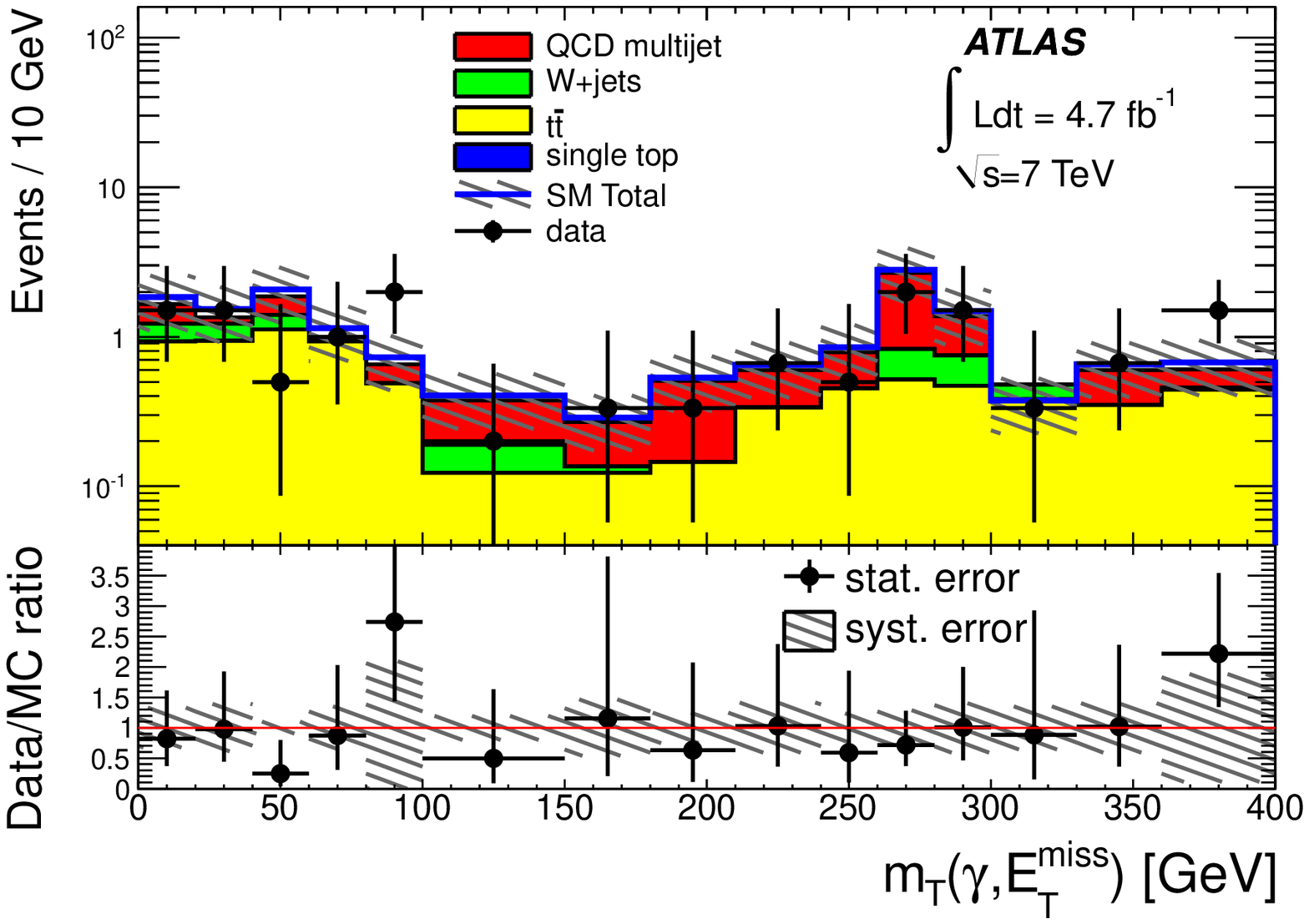}
\includegraphics*[width=0.49\textwidth]{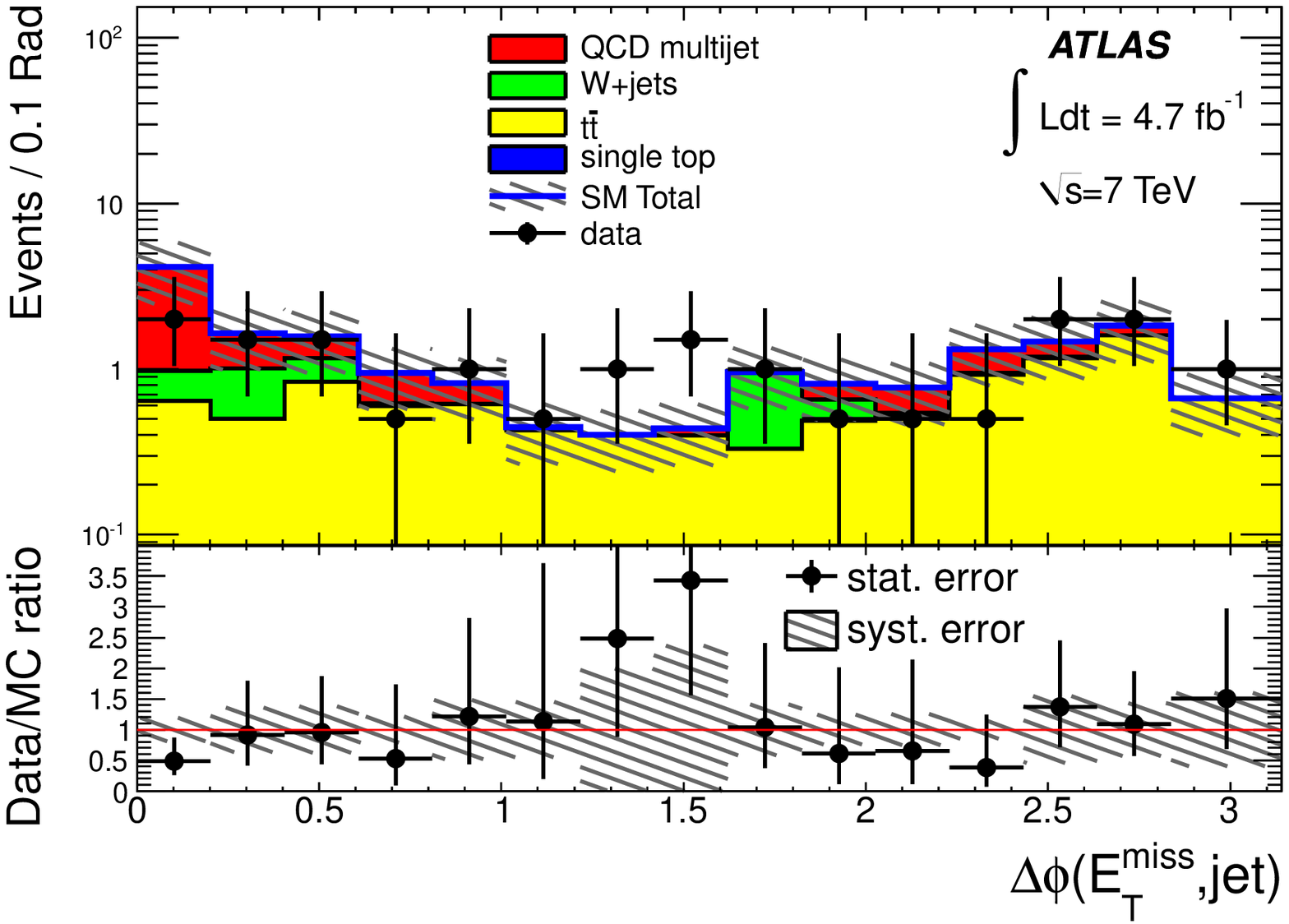}
\end{center}
\caption{The \mts~distribution after all selection criteria except the one for $\Delta\phi(\etmiss,\mathrm{jet})$ (top) and the $\Delta\phi(\etmiss,\mathrm{jet})$ distribution after all selection criteria except the one on \mts~(bottom).}
\label{fig:MT_metisol_MC_Data}
\end{figure}

Table~\ref{table:results} summarizes the expected number of Standard Model events and observed data events in the signal region.
A total of $7.5 \pm 2.2$ events are expected for the no-signal hypothesis while 7 events are observed.
The distribution of $E_{\rm{T}}^{\rm{miss}}$ after all requirements except on \etmiss~itself is shown in Fig.~\ref{fig:MET_ptgamma_MC_Data}, along with the distribution of $p_{\rm{T}}^{\gamma}$ after all requirements except those on \mts~and $\Delta\phi(\etmiss,\mathrm{jet})$.
The distribution of \mts~after all requirements except that on $\Delta\phi(\etmiss,\mathrm{jet})$ and the distribution of $\Delta\phi(\etmiss,\mathrm{jet})$ after all requirements except that on \mts~are shown in Fig.~\ref{fig:MT_metisol_MC_Data}.
The observed data agree with the background-only predictions.

\begin{table}[!ht]
\begin{center}
\begin{tabular}{lc}
\hline
Background source & Expected events  \tabularnewline
\hline
Electron misidentified as photon  & 1.1~$\pm$~0.1  \tabularnewline
Prompt photon and misidentified jet/$\tau$ & 3.4~$\pm$~1.8  \tabularnewline
QCD multijet & 2.7~$\pm$~1.1  \tabularnewline
$Z(\rarrow\nu\nu$)+jets & 0.3~$\pm$~0.3  \tabularnewline
\hline
Total number of expected events & 7.5~$\pm$~2.2  \tabularnewline
\hline
Observed events in the data & 7 \tabularnewline
\hline
  \end{tabular}
  \caption{Summary of the expected number of Standard Model events in the signal region and the number of events observed in the data. The systematic and statistical uncertainties, both included, are of the same order.}
 \label{table:results}
\end{center}
\end{table}

Since no excess is observed above the background-only prediction, the main result of the search is to constrain contributions from physics beyond the Standard Model.
The profile likelihood is used with an asymptotic approximation and the $\mathrm{CL_s}$ method to calculate confidence limits~\cite{Cowan:2010js, Read:2002hq}.
From the number of observed and expected events, a 95\% confidence level upper limit on the visible cross section, defined by the product of production cross section times efficiency times acceptance, is derived.
The expected 95\% confidence limit is 8.1 events, corresponding to an upper limit on the visible cross section of $1.7\,\mathrm{fb}$.
The observed limit is 7.4 events, corresponding to a visible cross section of $1.6\,\mathrm{fb}$.

\begin{figure}[!htbp]
\begin{center}
\includegraphics*[width=0.49\textwidth]{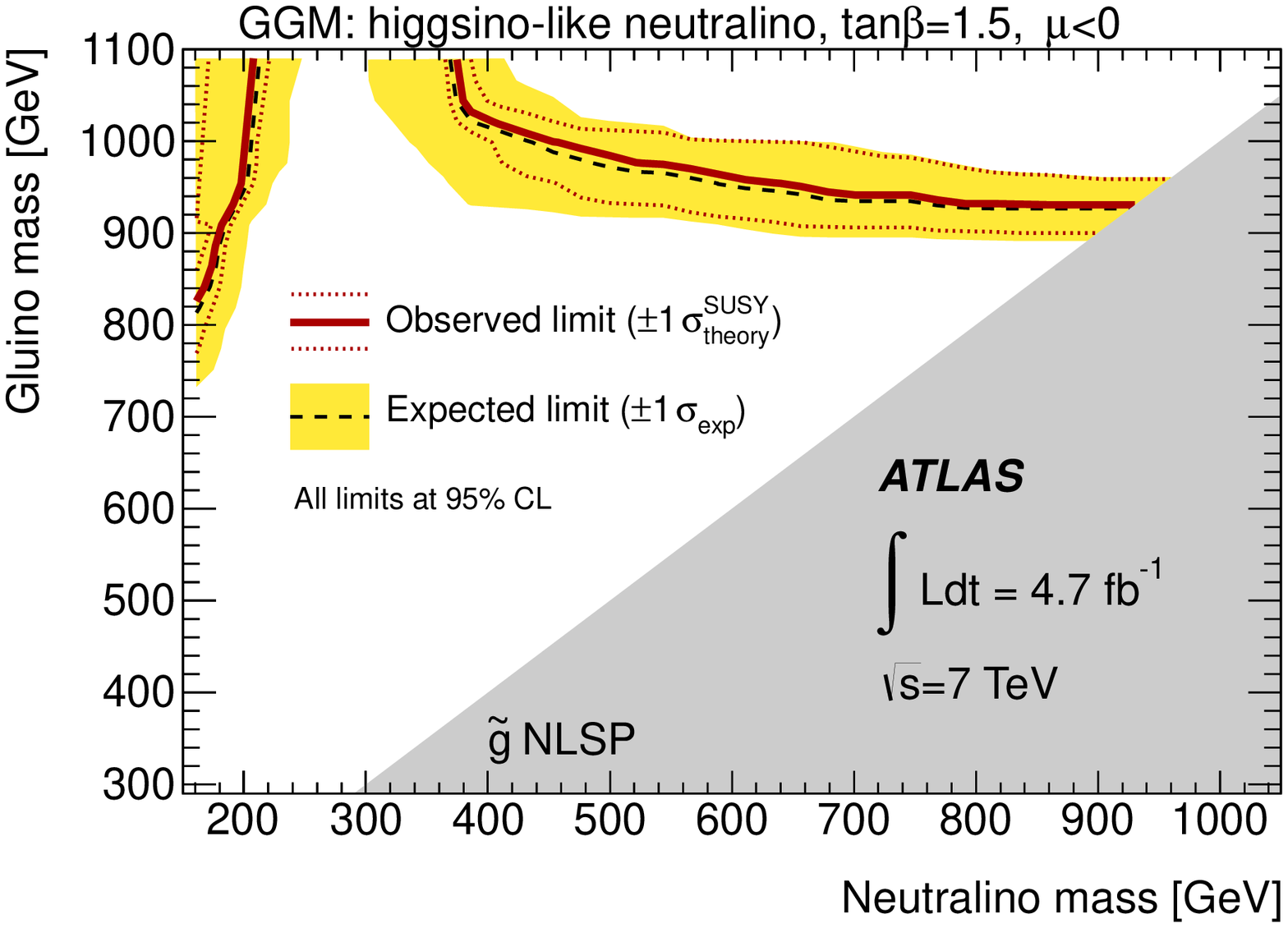}
\includegraphics*[width=0.49\textwidth]{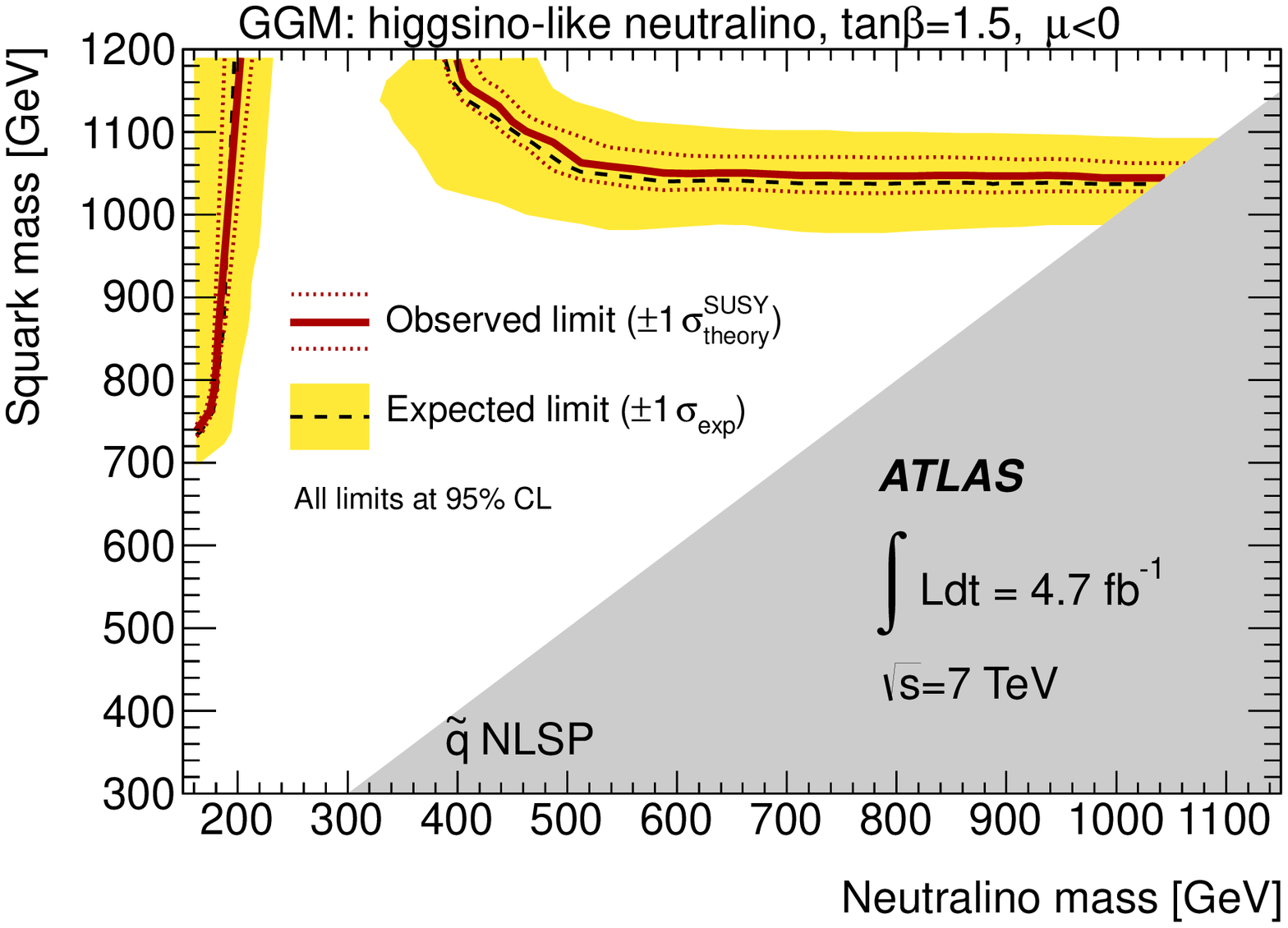}
\end{center}
\caption{Excluded region in the gluino-neutralino benchmark plane (top) and the squark-neutralino plane (bottom). The solid contour marks the observed exclusion, the dotted contours mark the observed exclusions if the signal cross section is shifted by $\pm 1\sigma$, and the dashed black line marks the expected exclusion. The shaded band gives the $\pm 1 \sigma$ ranges of the expected limit distribution.  The grey lower-right regions, corresponding to models with gluino or squark NLSP, are not considered.}
\label{fig:gl_neut_limits}
\end{figure}

The calculated acceptances for the simulated signal events and their cross sections are used in the framework of the specific GGM models described in Section~\ref{sec:intro} to map the excluded signal region.
For each point in the benchmark plane observed upper limits on the signal strength are calculated, including both strong production of squarks and gluinos and weak production of neutralinos and charginos.
Observed and expected limits for the combined production processes are shown in Fig.~\ref{fig:gl_neut_limits}.
The grey zone at the lower right represents the theoretically forbidden region where the lightest neutralino is no longer the NLSP.
Gluino masses less than $900\,\mathrm{GeV}$ and squark masses less than $1020\,\mathrm{GeV}$ for neutralino masses above $220\,\mathrm{GeV}$ are excluded in the respective planes.
The production cross sections at points with high neutralino mass and high gluino or squark mass are low, and an insufficient number of events is expected there.
For points at very low neutralino mass the cross section is high but the expected \etmiss~is relatively low and only a small fraction of events pass the event selection.
For this region the direct gaugino production through weak interactions is sizeable.
The points at intermediate neutralino mass $(220 < \mnt < 380\,\mathrm{GeV})$ are excluded by this search, regardless of gluino or squark masses, purely on the basis of the expected weak production. This gives rise to the ``chimney''-shaped exclusion region extending beyond the top edge of the benchmark plane.

If a Higgs boson mass $m_h=125\,\mathrm{GeV}$ is used instead of $115\,\mathrm{GeV}$, the branching ratio to $b\bar{b}$ is reduced, and the exclusion is weakened.
The important differences in excluded cross section for supersymmetric particle production, at high gluino mass and moderately high neutralino mass, are about 10\%.
In this relevant region, a 10\% change in cross section corresponds to a $10\,\mathrm{GeV}$ reduction in the $900\,\mathrm{GeV}$ gluino mass exclusion.

\section{Conclusions}\label{sec:conclusions}
A search for supersymmetry with a signature consisting of an isolated high transverse momentum photon, a $b$-tagged jet, and high missing transverse momentum is performed using $4.7\,\mathrm{fb}^{-1}$ of $\sqrt{s}$=7 TeV $pp$ collision data recorded with the ATLAS detector at the LHC.
Seven events are observed, consistent with the expected Standard Model background of $7.5 \pm 2.2$ events.
A model-independent 95$\%$ confidence level upper limit of $1.6\,\mathrm{fb}$ is set on the visible cross section of events passing the selection.
The cross-section limits are used to constrain higgsino-like neutralino production for a typical GGM model in two benchmark planes.
These are the first direct experimental constraints on this signature.
For neutralino masses greater than $220\,\mathrm{GeV}$, this search excludes gluino masses less than $900\,\mathrm{GeV}$ and squark masses less than $1020\,\mathrm{GeV}$ in the gluino-neutralino and squark-neutralino benchmark planes, respectively.



\section{Acknowledgements}

We thank CERN for the very successful operation of the LHC, as well as the
support staff from our institutions without whom ATLAS could not be
operated efficiently.

We acknowledge the support of ANPCyT, Argentina; YerPhI, Armenia; ARC,
Australia; BMWF and FWF, Austria; ANAS, Azerbaijan; SSTC, Belarus; CNPq and FAPESP,
Brazil; NSERC, NRC and CFI, Canada; CERN; CONICYT, Chile; CAS, MOST and NSFC,
China; COLCIENCIAS, Colombia; MSMT CR, MPO CR and VSC CR, Czech Republic;
DNRF, DNSRC and Lundbeck Foundation, Denmark; EPLANET, ERC and NSRF, European Union;
IN2P3-CNRS, CEA-DSM/IRFU, France; GNSF, Georgia; BMBF, DFG, HGF, MPG and AvH
Foundation, Germany; GSRT and NSRF, Greece; ISF, MINERVA, GIF, DIP and Benoziyo Center,
Israel; INFN, Italy; MEXT and JSPS, Japan; CNRST, Morocco; FOM and NWO,
Netherlands; BRF and RCN, Norway; MNiSW, Poland; GRICES and FCT, Portugal; MERYS
(MECTS), Romania; MES of Russia and ROSATOM, Russian Federation; JINR; MSTD,
Serbia; MSSR, Slovakia; ARRS and MVZT, Slovenia; DST/NRF, South Africa;
MICINN, Spain; SRC and Wallenberg Foundation, Sweden; SER, SNSF and Cantons of
Bern and Geneva, Switzerland; NSC, Taiwan; TAEK, Turkey; STFC, the Royal
Society and Leverhulme Trust, United Kingdom; DOE and NSF, United States of
America.

The crucial computing support from all WLCG partners is acknowledged
gratefully, in particular from CERN and the ATLAS Tier-1 facilities at
TRIUMF (Canada), NDGF (Denmark, Norway, Sweden), CC-IN2P3 (France),
KIT/GridKA (Germany), INFN-CNAF (Italy), NL-T1 (Netherlands), PIC (Spain),
ASGC (Taiwan), RAL (UK) and BNL (USA) and in the Tier-2 facilities
worldwide.


\bibliographystyle{model1a-num-names}
\bibliography{PhotonBMET}

%
\onecolumn
\clearpage
\begin{flushleft}
{\Large The ATLAS Collaboration}

\bigskip

G.~Aad$^{\rm 48}$,
T.~Abajyan$^{\rm 21}$,
B.~Abbott$^{\rm 111}$,
J.~Abdallah$^{\rm 12}$,
S.~Abdel~Khalek$^{\rm 115}$,
A.A.~Abdelalim$^{\rm 49}$,
O.~Abdinov$^{\rm 11}$,
R.~Aben$^{\rm 105}$,
B.~Abi$^{\rm 112}$,
M.~Abolins$^{\rm 88}$
% ...
\bigskip
\\
% ...
$^{11}$ Institute of Physics, Azerbaijan Academy of Sciences, Baku, Azerbaijan\\
$^{12}$ Institut de F{\'\i}sica d'Altes Energies and Departament de F{\'\i}sica de la Universitat Aut{\`o}noma de Barcelona and ICREA, Barcelona, Spain\\
% ...
$^{21}$ Physikalisches Institut, University of Bonn, Bonn, Germany\\
% ...
$^{48}$ Fakult{\"a}t f{\"u}r Mathematik und Physik, Albert-Ludwigs-Universit{\"a}t, Freiburg, Germany\\
$^{49}$ Section de Physique, Universit{\'e} de Gen{\`e}ve, Geneva, Switzerland\\
% ...
$^{88}$ Department of Physics and Astronomy, Michigan State University, East Lansing MI, United States of America\\
% ...
$^{105}$ Nikhef National Institute for Subatomic Physics and University of Amsterdam, Amsterdam, Netherlands\\
% ...
$^{111}$ Homer L. Dodge Department of Physics and Astronomy, University of Oklahoma, Norman OK, United States of America\\
$^{112}$ Department of Physics, Oklahoma State University, Stillwater OK, United States of America\\
% ...
$^{115}$ LAL, Universit{\'e} Paris-Sud and CNRS/IN2P3, Orsay, France\\
% ...
\end{flushleft}


\end{document}